# High-Frequency Domain Wall Oscillations in Ferromagnetic Nanowire with a Nanoscale Dzyaloshinskii–Moriya Interaction (DMI) Region

Durgesh Kumar[1], Rachid Sbiaa[2], Pinaki Sengupta[1], and S.N. Piramanayagam[1]

[1]School of Physical and Mathematical Sciences, Nanyang Technological University, 21 Nanyang Link,   Singapore 637371

[2]Department of Physics, Sultan Qaboos University, Muscat, 123, Oman

Corresponding Author: prem@ntu.edu.sg

The Dzyaloshinskii–Moriya Interaction (DMI) has laid the foundation for many novel chiral structures such as Skyrmions. In most of the studies so far, the DMI is present in the whole of the magnetic layer. Here, we report our investigations on a ferromagnetic nanowire where DMI is confined to a nanoscale region. We observe that the local modulation of magnetic properties causes oscillation of domain walls under the influence of spin-transfer torque. The oscillation frequency is tunable within a few GHz, making this observation potentially useful for applications in neuromorphic computing.

It has been almost fifty years since Dzyaloshinskii and Moriya have observed and explained the non-collinear interaction between spins [1,2]. However, only recently, the Dzyaloshinskii–Moriya Interaction (DMI) has picked up the interest of researchers [3-14]. Chiral domain walls that exhibit ultra high speed and Skyrmions are some highlights of DMI-induced phenomena [15-20]. In particular, for domain wall (DW) devices in spintronic applications, magnetic systems with interfacial DMI result in more rigid, stable and Chiral DWs, which move faster and more efficiently. Emori et al. studied current induced domain wall dynamics in Pt/CoFe/MgO and Ta/CoFe/MgO films and observed the domain wall motion due to Slonczewski-like torque from spin Hall effect [21]. The DMI from the interface results in the Nèel walls with a fixed chirality in place of Bloch walls and hence showed larger DW velocity. Ryu et al. deduced a similar conclusion for Pt/Co/Ni/Co/Pt films [20]. Yang et al. studied current driven DW dynamics in Co/Ni/Co/Ru/Co/Ni/Co/Pt synthetic antiferromagnetic (SAF) coupled structure, where chiral DWs were stabilised due to DMI from the Pt layer [22]. A higher DW velocity of  750 m/s was observed in this study. Since the DWs with the same chirality repel each other, they can be pushed synchronously along a nanowire. Kim et al. demonstrated the movement of several chiral DWs simultaneously upon the application of an out-of-plane pulsed magnetic field [23]. Memory or logic devices based on the above research activities are expected to revolutionize the electronics industry. On the fundamental front, investigation of DMI has attracted heightened interest because of its role in stabilizing several magnetic phases with non-collinear and non-coplanar spin orderings. Many of these phases exhibit exotic low energy excitations obeying fractional statistics and drive unconventional electronic transport phenomena such as linear magneto-resistance and topological Hall effect. These phenomena have been the subject of intense experimental and theoretical investigation over the past decade.

In most of the studies involving DMI so far, the investigations were focussed on a ferromagnetic layer which is entirely under DMI. We decided to investigate a distinctive scenario, where DMI acts only in a narrow region of a ferromagnetic nanowire. When a spin-polarized current is sent through the nanowire as in a domain wall device, a spin-transfer torque (STT) acts upon the magnetization. In this configuration, an interfacial DMI results in the tilting of spins in the ferromagnetic layer in one particular direction. When an electrical current flows along the nanowire, STT pushes the DW. However, the DW also experiences an additional restoring force in this DMI modified region. As a result, DW was found to oscillate with the microwave frequency for a limited range of applied current. In this report, we present our detailed micromagnetic simulation and explain this exciting phenomenon using an analytical model.

The simulation model describes a ferromagnetic (FM) nanowire, where Dzyaloshinskii–Moriya Interaction (DMI) is introduced in a nanoscale region. To systematically study this unusual configuration, we performed micromagnetic simulations using MuMax3 software, which uses the extended Landau-Lifshitz-Gilbert (LLG) equation for STT induced DW dynamics [24-27],

$$\frac{\partial \vec{m}}{\partial t} = -\gamma_0 \vec{m} \times \vec{H}_{eff} + \alpha \vec{m} \times \frac{\partial \vec{m}}{\partial t} - \vec{m} \times (\vec{m} \times [b\vec{J}.\vec{\nabla}]\vec{m}) + \beta \vec{m} \times [b\vec{J}.\vec{\nabla}]\vec{m} \qquad (1)$$

Here $\vec{m}$ is the locally reduced magnetization, $\vec{H}_{eff}$ is the effective magnetic field and $\alpha$ is Gilbert damping parameter. $\gamma_0 = \mu_0 \gamma$ where $\gamma$ is the gyromagnetic ratio, defined as $\gamma = \frac{ge}{2m_e}$, ($g$, $e$ and $m_e$ are Lande factor, electronic charge, and electronic mass, respectively). Moreover, $\vec{J}$ denotes the current density, $\beta$ is the degree of non-adiabaticity factor and $b = \frac{P\mu_B}{eM_s(1+\beta^2)}$. Here, $M_s$, $P$ and $\mu_B$ are saturation magnetization, the polarization of spin current and Bohr magneton, respectively. In equation (1), the first term represents field-induced precession of magnetization, and the second term represents magnetic damping, which tends to damp the precession. The last two terms represent the effects arising from an injected spin-polarized current. Among these, the first term describes the adiabatic while the second term describes the non-adiabatic torque from the applied electrical current to conduction electrons of ferromagnetic host nanowire.

The dimensions of the FM nanotrack were fixed to 300 nm × 30 nm × 7 nm, as shown in Fig. 1 (a). The DMI region was introduced in the middle of the FM nanowire. The length of DMI introduced region ($d$) was varied as 5 nm, 10 nm, 20 nm, 30 nm, 33 nm, 35 nm, 37 nm, and 40 nm. The DMI values used in the simulations were varied from -0.5 mJ/m$^2$ to -3 mJ/m$^2$ with a step of -0.5 mJ/m$^2$. The magnetic parameters such as saturation magnetization ($M_s$), exchange constant ($A$), anisotropy constant ($K_u$), damping constant ($\alpha$), non-adiabatic parameter ($\beta$) were chosen as 6.55×10$^5$ A/m, 1×10$^{-11}$ J/m, 4×10$^5$ J/m$^3$, 0.02, 0.02 respectively, which represent the parameters of (Co/Ni)$_n$ multilayer system [28-33]. Unless stated otherwise, these parameters are kept the same.

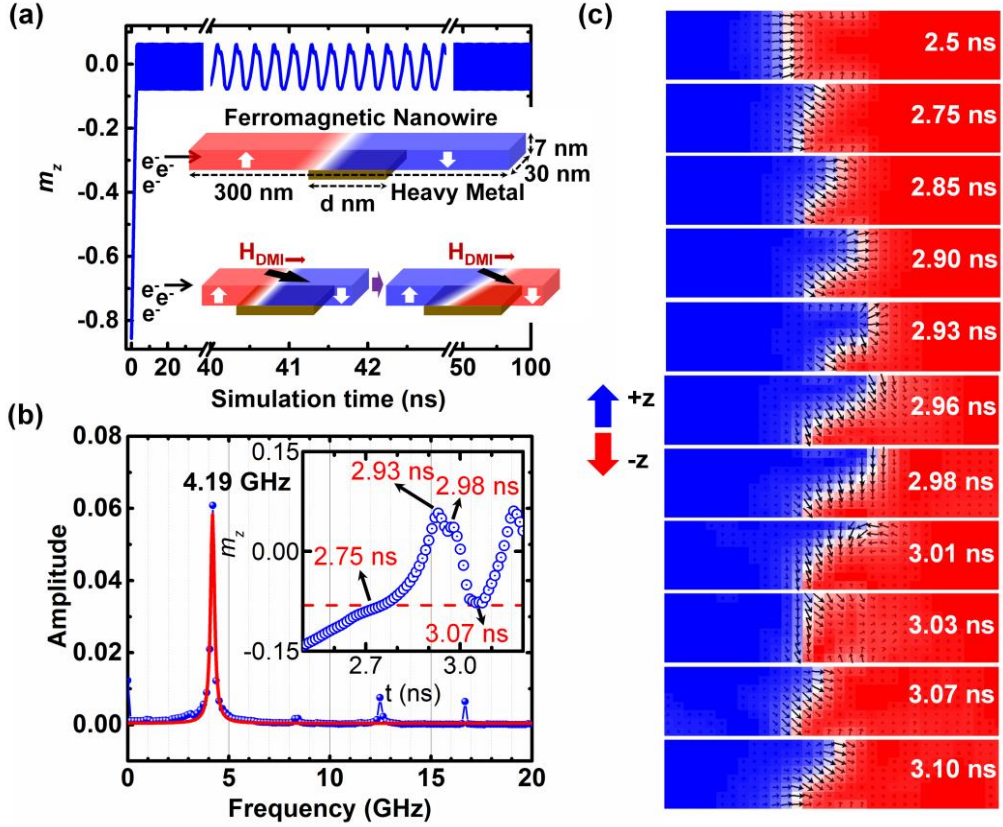

**Fig. 1.** (a) The variation of z-component of magnetization ($m_z$) as a function of simulation time ($t$) for d= 30 nm, DMI= -3 mJ/m$^2$ and J= 5×10$^{11}$ A/m$^2$. The magnetization oscillates with a certain frequency after the domain wall enters the magnetically modified region. Inset: the schematic of the proposed nanotrack and the effect of $H_{DMI}$ in the DW in subsequent times (b) the fast Fourier transform of Fig. 1 (a), which depicts the occurrence of first-order frequency peak at 4.19 GHz. Inset: zoomed in plot of $m_z$ vs t (simulation time) and (c) snapshots of the DW position at different simulation times.

Out of several ranges of parameters that we investigated, the DW was found to oscillate for values of d between 30 and 40 nm and DMI strength between -2.5 and -3 mJ/m$^2$. The variation of *z*-component of magnetization ($m_z$) as a function of simulation time ($t$) was used to describe the DW oscillation. Fig. 1 (a) shows the DW oscillation for the nanowire with *d* of 30 nm, DMI value of -3 mJ/m$^2$ and a current density (*J*) of 5×10$^{11}$ A/m$^2$. Once the DW reaches the DMI region, the lower end of the DW gets pinned at the lower edge of the DMI region, and the other end of the DW oscillates back and forth with a certain frequency. The frequency of the oscillations was found to be 4.19 GHz (shown in Fig. 1 (b)), as obtained from the fast Fourier transform (FFT) of Fig. 1 (a) [34]. Figure 1 (c) depicts the snapshots of DW positions for different simulations times. The blue and red colors represent the magnetization along +z and -z directions, respectively. As can be seen, the DW enters in the DMI modified region at *t* of 2.75 ns and reaches the other end at *t*~ 2.93 ns. At *t*~ 3.03 ns, the DW moves back and $m_z$ attains the same value as it was at $t= 2.75$ ns. This cycle repeats for subsequent simulation time (see supplementary for video).

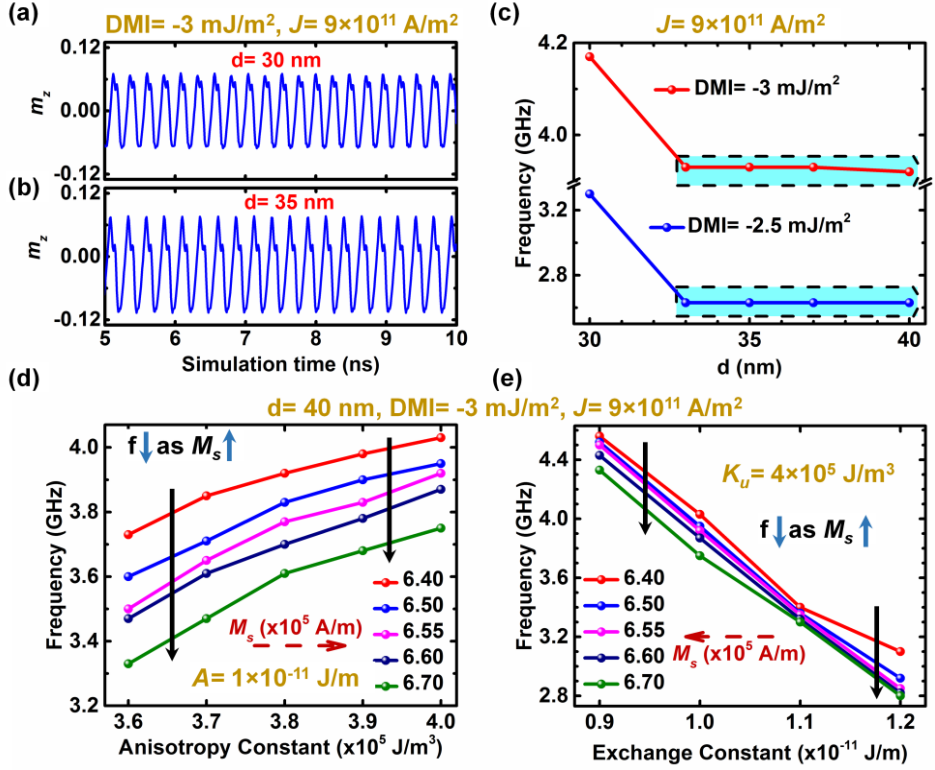

**Fig 2**. DW oscillation (i.e. $m_z$ vs $t$) for (a) d= 30 nm, (b) d= 35 nm at DMI at -3 mJ/m² and (c) Plot of frequency of DW oscillations for different d values. The dependence of frequency of oscillations on (d) anisotropy constant ($K_u$), and (e) exchange constant ($A$) for various values of saturation magnetization ($M_s$).

In order to understand the effect of material and design parameters on the oscillatory behavior of the DW, we carried out detailed investigations and the results are summarized in Fig. 2. Fig. 2 (a) and (b) show the oscillation profile of DW for $d = 30$ and 40 nm, respectively and Fig. 2 (c) shows the frequency for different values of $d$ ranging from 30 to 40 nm and DMI strength of -2.5 and -3 mJ/m². Except for $d= 30$ nm, the frequency is found to be independent of $d$ for both investigated cases, indicating that the observed behavior may not have a geometrical origin. It should be noticed that the oscillation frequency increases with the current density. We performed a full set of simulations for many different cases. Based on these simulation results, we fixed the current density at $9 \times 10^{11}$ A/m² for simulations in Fig. 2 and 3 (see supplementary).

Since the current induced DW dynamics strongly depends on the magnetic parameters of the material, we also investigated the dependence of oscillation frequency on these parameters such as $K_u$, $A$ and $M_s$. Fig. 2 (d) and 2 (e) summarizes the dependence of frequency on these parameters. As $K_u$ increases, the frequency increases with the maximum observed frequency of 4.03 GHz in this set of simulations. However, the frequency decreases as $A$ increases. In this set of simulations, we observed a maximum frequency of 4.56 GHz. In both cases, the frequency decreases as $M_s$ increases. By tuning $A$, $K_u$ and $M_s$, we observed a maximum frequency of 4.97 GHz (see suplementary).

The results of the dependence of oscillation frequency on $M_s$, $A$ and $K_u$ follow the micromagnetism for current induced DW dynamics (Eq. (1)), but at the same time, they are also affected by the DMI field. In the experimental realization of this concept, a heavy metal (HM) with a high spin-orbit coupling may be deposited on the top or the bottom of the FM layer. The structural inversion asymmetry (SIA) would lead to an interfacial DMI in the ferromagnetic layer. As a result, the form of $H_{eff}$ in Eq. (1) gets modified with an extra term, as below [35,36]

$$\vec{H}_{DMI} = \frac{2D}{\mu_0 M_s}\left(\frac{\partial m_z}{\partial x}, \frac{\partial m_z}{\partial y}, -\frac{\partial m_x}{\partial x} - \frac{\partial m_y}{\partial y}\right) \qquad (2)$$

Where $D$ is the DMI parameter. The magnitude of the DMI-induced contribution to the effective magnetic field is [3,37,38],

$$H_{DMI} = \frac{2D}{\mu_0 M_s}\left(\frac{\sqrt{K_u - \left(\frac{\mu_0 M_s^2}{2}\right)}}{\sqrt{A}}\right) \qquad (3)$$

The presence of DMI causes a significant effect on DW statics as well as dynamics. The first impact of DMI is that the structure of the DW changes. With increasing DMI, the structure changes from pure Bloch DW to mixed Bloch and Nèel DW, and beyond a critical value, the DW stabilizes in the pure Nèel DW structure. In other words, DMI acts similar to an effective field for DW in a direction normal to the DW, and this effective field rotates the DW magnetization pointing in a direction from up-domain to down-domain. Besides the change in the DW structure which is usually observed in every study, the nano-scale DMI leads to a tilting of the DW surface. This tilting was observed to be proportional to the strength of DMI. Thus, when the DW enters the DMI active region, there is an extra instantaneous force on the DW, which is proportional to $H_{DMI}$. Under suitable parameters and conditions, DW oscillation occurs in this region.

In this study, the DMI is introduced in a small portion of the ferromagnetic nanowire. Therefore, when the DW enters the DMI active region, the structure of the DW changes but it may not reach a steady (or minimum energy) state. In addition, the DW tilts in this region, and the lower edge of the DW gets pinned to the beginning of the (DMI introduced) pinning region. The upper end moves following the tilting of DW. As the upper end slightly crosses the pinning region (the other end of DMI active region), the DW experiences a different energy barrier (or torque acting on the DW) as there is no DMI. If the torque from STT is not enough to depin the lower edge from the lower end of pinning site (starting point of DMI active region), the DW comes back, and this process continues and results in oscillations. It is important to note that the DW moves in the +x direction because of the presence of STT and not because of the existence of DMI.

In order to understand the results in detail, we have modeled the dynamics of the DW with that of a single spin (spin at the center of DW) in a Zeeman field (analogous to $H_{DMI}$). Figure 3 (a) shows the schematic of the analytical model. The energy of the spin in this Zeeman-like field is given by [39,40],

$$U = -\mu_0(\vec{m} \cdot \vec{H}_{DMI}) \quad (4)$$

$$U = -\mu_0\, m\, H_{DMI} \cos\theta \quad (5)$$

The DMI does not change the magnitude of the spin, but only changes the angular orientation of the spin. The magnitude of force acting on this spin becomes,

$$F = -(\mu_0\, m\, H_{DMI} \sin\theta) \quad (6)$$

For small $\theta$, Eq. (6) can be rewritten as,

$$F = -(\mu_0\, m\, H_{DMI}\, \theta) \quad (7)$$

which is the equation of motion for classical harmonic oscillator and $H_{DMI}$ acts as a restoring force for DW oscillations. Therefore, the frequency of oscillation is given by,

$$\omega \propto \sqrt{H_{DMI}} \text{ (with } \omega = 2\pi f) \quad (8)$$

Since STT drags the DW in the +x direction (in our case), the force due to STT is in the +x direction. At point "a", the force due to DMI, $F_{DMI}= 0$ (since $\theta= 0$, from Eq. (6)). Therefore, $F_{STT}$ will drag the DW further, but with further movement, $\theta$ will be non-zero. Consequently, there will be a force $F_{DMI}$ in the negative x-direction, which increases with $\theta$. Hence, the DW will oscillate between the boundaries of these two forces.

From the analytical model, it is clear that the frequency depends on $H_{DMI}$. Since $H_{DMI}$ depends on $A$, $K_u$ and $M_s$, the results in Fig. 2 can be readily explained using Eqs. (3) and (8). From Eq. (8), it is also imperative that the frequency must depend on $\sqrt{D}$. To verify this relationship, we carried out simulations for various values of DMI strengths. Figure 3(b) depicts the dependence of f on D. Except for the region between 2.5 and 2.6 mJ/m$^2$, where the onset of oscillation occurs, the rest of the region follows a D$^{1/2}$ dependence.

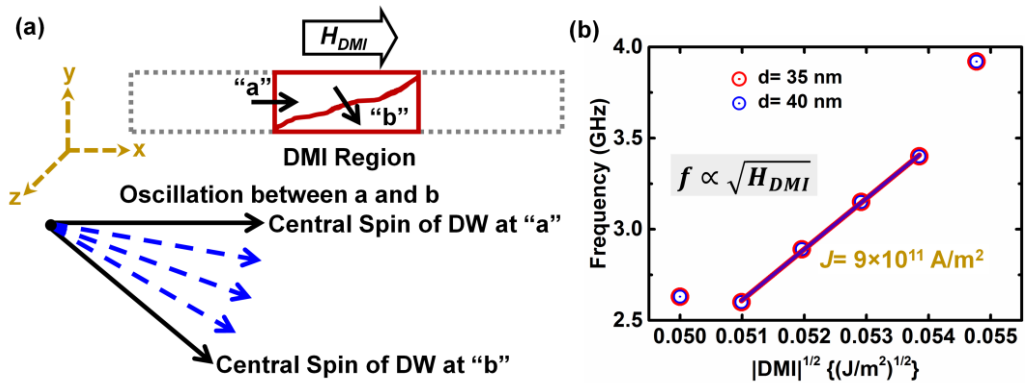

**Fig. 3.** (a) Schematic of the nature of DW oscillation for an analytical explanation and (b) the dependence of the frequency of oscillations with DMI.

Based on the results discussed above, we propose a device as shown in Fig. 4. The device is composed of a ferromagnetic nanowire with a heavy metal deposited (on either top or bottom of the nanowire) in a portion where we wish to introduce magnetic modification through DMI. The heavy metal can be arranged in a crossbar configuration. In the opposite side of the HM crossbar, a tunnel barrier and a reference ferromagnetic layer are deposited. The magnetization of the reference layer will be fixed and following the DW oscillation, a change in the relative magnetization direction of these two layers will occur, which will be measured by the change in tunneling magnetoresistance (TMR). This change in TMR can be converted to a microwave signal. Thus, this device converts a DC voltage into AC voltage at microwave frequencies. Since, the frequency of oscillation can be tuned from a minimum of 1.82 GHz to a maximum of 4.97 GHz by changing the width of DMI introduced region ($d$), value of DMI, current density ($J$) and magnetic parameters such as $M_s$, $A$, $K_u$ etc., a number of such oscillators can be fabricated in a single chip. Thus, a range of frequencies can be obtained. Such an array of oscillators are useful in recently proposed spin-based neuromorphic computing.

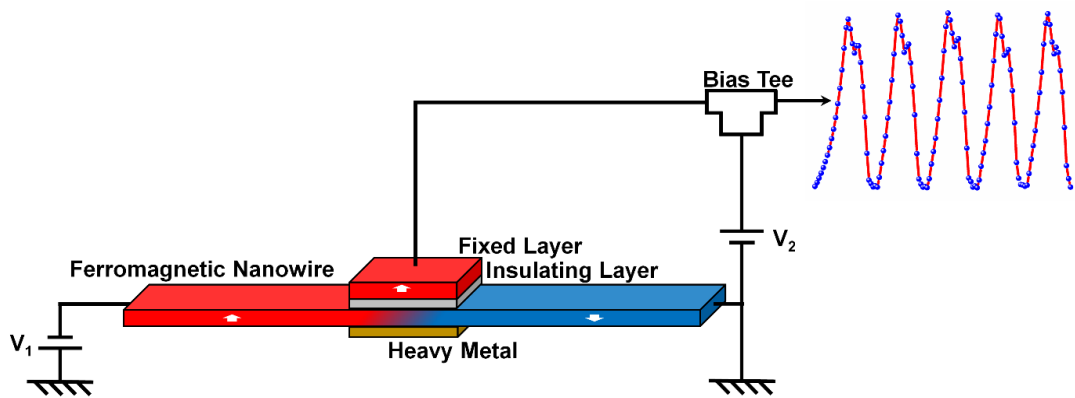

**Fig. 4.** The schematic of the proposed oscillator with the HM crossbar at one side and MTJ at the other side of the ferromagnetic nanowire.

In this work, we have shown that for a finite range of $d$, DMI, and $J$, the DW oscillates at the magnetically modified region through DMI. The DMI results in an effective field, which acts on the DW locally. In addition to STT from applied DC current, an extra force from DMI makes the DW oscillating for certain values of geometrical and magnetic parameters. This observation leads to a device which can generate a wide range of frequencies from a single chip, useful for neuromorphic computing applications.